\documentstyle[preprint,revtex]{aps}
\def\be{\begin{equation}}
\def\ee{\end{equation}}
\begin{document}
\draft
\preprint{IFT-P.040/92}
\preprint{IFUSP/P-1023/92}
\preprint{December 1992}
\begin{title}
 Constraints on singlet right-handed neutrinos  \\
coming from the $Z^0$-width
\end{title}
\author{C. O. Escobar$^{a}$, O.L.G. Peres$^{b}$, V. Pleitez$^{b}$
\\and \\R. Zukanovich Funchal$^{a}$ }
\begin{instit}
$^{a}$ Instituto de F\'\i sica da Universidade de S\~ao Paulo\\
01498-970 C.P. 20516--S\~ao Paulo, SP\\
Brazil\\
$^{b}$Instituto de F\'\i sica Te\'orica \\
Universidade Estadual Paulista \\
Rua Pamplona, 145 \\
01405-900--S\~ao Paulo, SP \\
Brazil
\end{instit}
\begin{abstract}
We study the constraints on masses and mixing angles imposed by the measured
$Z^0$ invisible width, in a model in which a singlet right-handed neutrino
mixes with all the Standard Model  neutrinos.
\end{abstract}
If neutrinos are massive an important question to be answered
concerns the way the $Z$-pole observables constraint their masses and
mixing parameters. In particular the measured $Z$-invisible width,
$\Gamma^{inv}$, implies that the number of families is
compatible with three. On the other hand, it is well known that
this number need not to be an integer number if right-handed
neutrinos transforming as singlets under $SU(2)_L\otimes U(1)_Y$ are
added to the particle content of the theory.

Experimental searches for sequential neutral leptons beyond the three
generations exclude stable
Dirac neutrinos below $41.8$ GeV and stable Majorana neutrinos below
$34.8$ GeV. For the unstable case these values are $46.4$ and $45.1$
GeV respectively~\cite{l3}. However, it is worth stressing that these
limits are valid for sequential leptons and do not apply to the
case of singlets of right-handed neutrinos.

Here we will consider the simplest extension of the standard
electroweak model~\cite{ws} with the addition of one right-handed
singlet neutral fermion~\cite{cj}, resulting in 4 physical neutrinos
two of them massless and two massive ones ($\nu_P,\nu_F$). It has
been argued that this simple extension allows the implementation of a
see-saw mechanism~\cite{see} which would explain the smallness of the
known neutrinos masses. Hence, from the theoretical point of view,
usually this mechanism is assumed. In fact, within this view it has
been claimed that three generation and a single right-handed singlet
are ruled out by experiments ($Z$-invisible width and $\tau$
decay)~\cite{c1} and for this reason a fourth generation must be
added to the Standard Model besides the singlet~\cite{c2}.
However, one should not be limited by the see-saw
framework, as ultimately the question of neutrino masses will be
settled by experiments.

In fact in this letter three mass regions
will be considered for the massive neutrinos: (i) $m_P, m_F < M_Z/2$,
(ii) $m_P<M_Z/2$ and $M_Z/2<m_F<M_Z$ and (iii) $m_P<M_Z/2$,
$m_F>M_Z$.

We will take into account current experimental results for
the $Z^0$ invisible width in the particular model of Ref.~\cite{cj}.
Let us first briefly review the model we work with.

The matter fields are those in the Standard Model plus a singlet
right-handed neutrino. In this simple extension of the Standard
Model the most general form of the neutrino mass term is
\be
{\cal L}_{\nu}^{M}
=-\sum_{i=1}^{3}a_j\bar\nu'_{jL}\nu'_R-\frac{1}{2}M\overline{\nu'^c_R}
\nu'_R+H.c.
\label{mass}
\ee
Where the primed fields are not yet the physical ones.
In this model, there are four physical neutrinos
$\nu_1,\nu_2,\nu_P$ and $\nu_F$, the first two are massless and the last two
are massive Majorana neutrinos with masses
\be m_P=\frac{1}{2}(\sqrt{M^2+4a^2}-M),
\quad
m_F=\frac{1}{2}(\sqrt{M^2+4a^2}+M),
\label{mass2}
\ee
where $a^2=a_1^2+a_2^2+a_3^2$.
In terms of the physical fields the charged currents interactions are
\be
{\cal L}^{CC}=
\begin{array}{cccc}(\bar\nu_1& \bar\nu_2& \bar\nu_P&
\bar\nu_F\end{array})_L\gamma^\mu \Phi R
\left(\begin{array}{c}
e\\ \mu\\ \tau \\ 0 \end{array}\right)_LW^+_\mu+H.c.,
\label{cc}
\ee
where $\Phi=diag(1,1,i,1)$ and $R$ is the matrix
\be
\left(
\begin{array}{cccc}
c_\beta &-s_\beta s_\gamma & -s_\beta c_\gamma & 0 \\
0 & c_\gamma & -s_\gamma & 0 \\
c_\alpha s_\beta\; & c_\alpha c_\beta s_\gamma\; &c_\alpha c_\beta
c_\gamma\; &-s_\alpha \\
s_\alpha s_\beta \;& s_\alpha c_\beta s_\gamma\; & s_\alpha c_\beta
c_\gamma\;  & c_\alpha
\end{array}\right)
\label{mm}
\ee
In Eq.~(\ref{mm}) $c$ and $s$ denote the cosine and the sine of the
respective arguments. The angles $\alpha,\beta$ and $\gamma$ lie in
the first quadrant and are related to the mass parameter as follows
\be
s_\alpha=\sqrt{m_P/(m_P+m_F)},
\label{angles}
\ee
\be
s_\beta=a_1/a,\;c_\beta s_\gamma=a_2/a,\;c_\beta c_\gamma=a_3/a.
\label{angles2}
\ee
Notice that there are four parameters, we will choose two angles and the
two Majorana masses.

The neutral currents interactions  written in the physical basis read
\be
{\cal L}^{NC}=\left(\begin{array}{cccc}
\bar\nu_{1}&\bar\nu_{2}&\bar\nu_{P}&\bar\nu_{F}\end{array}\right)_L
\gamma^\mu\left(\begin{array}{cccc}
1\; \;& 0\; & 0 & 0 \\
0 \;\;& 1\; & 0 & 0 \\
0 \;\;& 0 \;& c^2_\alpha & ic_\alpha s_\alpha \\
0 \;\;& 0 \;& -ic_\alpha s_\alpha & s^2_\alpha\end{array}\right)
\left(\begin{array}{c}
\nu_{1}\\ \nu_{2} \\ \nu_{P} \\ \nu_{F}\end{array}\right)_L Z_\mu+H.c.
\label{nc}
\ee
The Majorana neutrinos have the property that their diagonal vector and
non-diagonal axial-vector couplings to the $Z^0$ vanish.

For the masses in region (i), both massive neutrinos contribute to the
invisible width of the $Z^0$ which is given by~\cite{cj}
\be
\Gamma^{inv}(Z\rightarrow
\nu's)=\Gamma_0 (2+c^4_\alpha \chi_{PP}+2c^2_\alpha s^2_\alpha \chi_{PF}
+s^4_\alpha \chi_{FF}),
\label{width}
\ee
where $\Gamma_0$ is the width for each massless neutrino pair in the Standard
Model.
\be
\chi_{ij}=\frac{\sqrt{\lambda (M^2_Z,m^2_i,m^2_j)}}{M^2_Z} X_{ij},
\label{chi}
\ee
$\lambda$ is the usual triangular function defined by
\[\lambda(a,b,c)=a^2+b^2+c^2-2(ab+ac+bc).\]
and $X_{ij}$ include the mass dependence of the matrix elements.
Explicitly,
\[X_{PP}=1-4\frac{m^2_P}{M^2_Z} \quad
X_{FF}=1-4\frac{m^2_F}{M^2_Z} \]
\be
X_{FP}=1-\frac{\Delta m^2_{FP}}{2M^2_Z}-\frac{m^2_P+3m_Fm_P}{M^2_Z}
-\frac{(\Delta m^2_{FP})^2}{4M^4_Z}
\label{x}
\ee
where we have defined $\Delta m^2_{FP}=m^2_F-m^2_P$.
Thus, $\chi_{ij}$ are bounded by unity whereby
\be
\Gamma^{inv}(Z\rightarrow \nu's)\leq 3\Gamma_0.
\label{width2}
\ee

The number of neutrinos $N_\nu$, defined as
$N_\nu=\Gamma^{inv}/\Gamma_0$, as determined from recent LEP
data~\cite{pdg} is
\be
N_\nu=3.00\pm0.05.
\label{n}
\ee
Next, it is interesting to see how this value of $N_\nu$
constrains the allowed values for the neutrino masses $m_P,m_F$.

Let us isolate the non-standard contribution in Eq.(8), and restrain
it to a region compatible with (12). With the help of
Eqs.~(\ref{angles}), (\ref{chi}) and (\ref{x}) we can write the
relation
\be
\frac{1}{(x+y)^2}\left[x^2F(y)+y^2F(x)+2xyG(x,y)\right]=1.00\pm0.05,
\label{f}
\ee
where we have defined
\[
F(\zeta)=(1-4\zeta^2)^{\frac{3}{2}},
\]
\be
G(x,y)=\sqrt{1+(x^2-y^2)^2-2(x^2+y^2)}\left[1-\frac{x^2+y^2}{2}
-3xy-\frac{(x^2-y^2)^2}{4} \right]
\label{def}
\ee
and we have used $x=m_F/M_Z$ and $y=m_P/M_Z$.

We show in Fig.\ref{f1} the contour plot of the function on the left-hand
side of Eq.(\ref{f}). As can be seen from Eq.
(\ref{mass2}) one of the neutrinos must be heavier than the other. We
choose the $M$ parameter to be positive, i.e. $m_F>m_P$. This limit
appears in Fig.\ref{f1} as the region below the $45^o$ straight line.
Notice from Eq.(\ref{width2}) that in this type of model the neutrino
counting will result in a number at most 3 even in the case, as here,
of 3 families~\cite{cj}. As a result the right-hand side of Eq.(13)
should be limited from above by $1.00$.

A blow-up of Fig.\ref{f1} is shown in Fig.\ref{f2} highlighting the
one and two standard deviation contours below the central value
$1.00$. From Eq.~(\ref{angles}) for the mixing angle $\alpha$, we
obtain $y=x\tan^2\alpha$. This straight line has to lie below the
before mentioned contours in order to be consistent with experiments.
This implies a constraint on the mixing angle $\alpha$, $\tan^2\alpha<0.055$,
i.e. $\sin\alpha<0.23$. In terms of the masses we have
\be
m_F>18.2\,m_P.
\label{mass3}
\ee
In this model the mixing in the charged current are as shown in
Eqs.(\ref{cc}) and (\ref{mm}).
{}From those couplings we see that $\Gamma(\tau\to\nu_P e^-\bar\nu_e)
\propto c^2_\alpha$ and $\Gamma(\tau\to\nu_F e^-\bar\nu_e)\propto
s^2_\alpha$. It has been pointed out that in order to solve the
puzzle of the leptonic $\tau$ decays $\sin\alpha$ must be between
$0.1$ and $0.3$~\cite{foot}. Although this is compatible with our
result we stress that as the two neutrinos can contribute to the
$\tau$ decay a more careful analysis must be done taking into
account the respective phase space factors~\cite{eppz}.

In the case of region (ii) all the masses in the kinematical region
are allowed. For the third region (iii), we have found that the
lightest neutrino mass has to be smaller than $9\,\mbox{GeV}$
which of course is not in conflict with the $\nu_\tau$ mass limit.

In conclusion we see that singlet right-handed neutrinos need not
necessarily come from a see-saw mechanism. In particular the region
(i) has a rich phenomenology which deserves to be studied~\cite{eppz}.

\acknowledgments
We are very grateful to Funda\c c\~ao de Amparo \`a Pesquisa do Estado de
S\~ao Paulo (FAPESP) (O.L.G.P. and R.Z.), for full financial support and
Con\-se\-lho Na\-cio\-nal de De\-sen\-vol\-vi\-men\-to Cien\-t\'\i
\-fi\-co e Tec\-no\-l\'o\-gi\-co (CNPq) (V.P.) for
partial financial support.

\figure{Contour plot of function (\ref{f}) where $x=m_F/M_Z$ and
$y=m_P/M_Z$. The $45^o$ straight line limits the region where
$m_F>m_P$.\label{f1}}
\figure{This blow-up of Fig.~\ref{f1} shows the regions constrained
by the measured $Z^0$ invisible width within one and two standard
deviation $(1-0.05,\,1-0.10)$.\label{f2}}
\end{document}